# Migration Networks: Applications of Network Analysis to Macroscale Migration Patterns[1]

Valentin Danchev*, Stanford University, USA

Mason A. Porter, University of California, Los Angeles, USA

## Abstract

An emerging area of research is the study of macroscale migration patterns as a network of nodes that represent places (e.g., countries, cities, and rural areas) and edges that encode migration ties that connect those places. In this chapter, we first review advances in the study of migration networks and recent work that has employed network analysis to examine such networks at different geographical scales. In our discussion, we focus in particular on global-scale migration networks. We then propose ways to leverage network analysis in concert with digital technologies and online geolocated data to examine the structure and dynamics of migration networks. The implementation of such approaches for studying migration networks faces many challenges, including ethical ones, methodological ones, socio-technological ones (e.g., data availability and reuse), and research reproducibility. We detail these challenges, and we then consider possible ways of linking digital geolocated data to administrative and survey data as a way of harnessing new technologies to construct increasingly realistic migration networks (e.g., using multiplex networks). We also briefly discuss new methods (e.g., multilayer network analysis) in network analysis and adjacent fields (e.g., machine learning) that can help advance understanding of macroscale patterns of migration.

## Keywords

---

[1] In this chapter, we draw on Danchev (2015) and Danchev and Porter (2018).

* vdanchev@stanford.edu

# Introduction

The complexity of international-migration patterns has increased markedly over the last few decades (International Organization for Migration, 2003). Advancements in distance-shrinking technologies of transportation, information, and communication (Harvey, 1989) have interacted with uneven global development (Wallerstein, 1974); migration policy restrictions (Hatton and Williamson, 2005); and international exchanges of goods, capital, and services (Sassen, 2007, Sassen, 1988, Held et al., 1999) to produce complex migration patterns between previously disconnected places across the globe. To account for such tendencies, migration theories (Salt, 2013 [1986], Kritz et al., 1992, Portes and Böröcz, 1989) have advanced the perspective that international migration is not merely a response to bilateral origin–destination forces, but also often takes place in a group (or in a network) of origins and destinations that are interconnected through prior migration, trade, culture, and politics. Building on this idea, one can study migration as a network of nodes (which typically represent countries but may alternatively represent rural areas, cities, provinces, or other geographically dispersed places) and edges (which encode migration ties that are based on movements of diverse populations, such as displaced individuals, low-skilled workers, or skilled professionals) between places. Such "migration networks" provide a useful perspective for the understanding of broader patterns of international migration.

In this chapter, we discuss how to leverage network analysis in concert with new technologies and data sources to study migration networks at several geographical scales. We focus in particular on migration networks at global scales. We first review recent work that has employed data from origin–destination matrices of bilateral migrant stock (Özden et al., 2011, UN DESA, 2019) and migration flows (Abel and Sander, 2014, Abel, 2018) to study various aspects of the structure and evolution of the global migration network[2] (Fagiolo and Mastrorillo, 2013, Davis et al., 2013, Tranos et al., 2015, Danchev and Porter, 2018, Windzio, 2018). We then discuss future research opportunities and ethical, methodological, data-availability, and reproducibility challenges that are associated with the use of digital geolocated data from online social networks, World Wide Web, and mobile-phone networks (Lazer et al., 2009, Salganik, 2018) to study migration networks. We consider possible ways of linking digital geolocated data to administrative and survey data as a way of constructing and studying increasingly realistic migration networks. We also review recent advancements in network methodology and modeling that can help advance network analysis in the field of migration.

---

[2] By convention, we refer to "the" global migration network as a singular entity. However, there are many possible realizations of a global migration network, depending on data sources and on the ways in which one constructs such a network.

Network analysis is a powerful tool for studying relationships between entities (e.g., individuals, countries, Web pages, computers, or neurons), rather than focusing solely on the entities themselves (Wasserman and Faust, 1994, Newman, 2018, Butts, 2009, Brandes et al., 2013). Scholars from diverse fields—including sociology (Wasserman and Faust, 1994), economics (Jackson, 2008), political science (Maoz, 2011), applied mathematics (Porter, 2020), statistics (Kolaczyk, 2009), physics (Newman, 2018), social neuroscience (Baek et al., in press (2020)), and others—have employed network analysis to study complex systems of interconnected entities. Taking into account system specificity, one can first carefully abstract a set of entities as a network's nodes and a set of relationships as edges (e.g., encoding social ties, migration ties, or hyperlinks) between those nodes (Butts, 2009, Brandes et al., 2013). One can then study patterns of relationships—in the form of network structure—that emerge from the interacting entities and examine how network structure affects the dynamics (and function) of a system and the features (e.g., importances or performance) of particular nodes, edges, and other substructures (Wasserman and Faust, 1994: 3, Borgatti et al., 2009: 894, Newman, 2018, Kolaczyk and Csárdi, 2014, Luke, 2015).

It is convenient to categorize applications of network analysis of migration into two strands (Cushing and Poot, 2003, Maier and Vyborny, 2008, Bilecen et al., 2018, Lemercier, 2010). The first strand focuses on *microscale networks, such as in the form of migrants' personal networks*, which one constructs based on interpersonal relationships (e.g., kinship, friendship, or acquaintance) that link migrants and non-migrants (Massey et al., 1998: 42, Boyd, 1989). Migrants' personal networks can spread information about destination opportunities and provide initial employment, accommodation, and overall assistance, thereby reducing movement costs and risks (MacDonald and MacDonald, 1964, Gurak and Caces, 1992, Massey et al., 1998, Boyd, 1989, Palloni et al., 2001, Liu, 2013, Blumenstock et al., 2019). For example, recent work has used data from interviews and surveys (Lubbers et al., 2010, Vacca et al., 2018, Verdery et al., 2018, Martén et al., 2019, Bilecen et al., 2018) and the online social network Facebook (Herdağdelen et al., 2016) and employed what is often called "social network analysis" (SNA) (Wasserman and Faust, 1994, Borgatti et al., 2018) to examine migrants' personal networks and the relative importances of migrants' ties to other migrants and their ties to non-migrants in home and host societies. The second strand, which uses aggregate data and focuses on *macroscale migration networks*, encompasses a multidisciplinary body of literature that leverages network methodology to study migration as a "mechanism that connects 'places'" (Maier and Vyborny, 2008). Among other phenomena, researchers have examined networks of internal (intra-country) movements that connect rural areas in Northern France (Lemercier and Rosental, 2010), movements that connect different states in the United States (Maier and Vyborny, 2008, Charyyev and Gunes, 2019), and global networks of international-migration ties that connect different countries (Fagiolo and Mastrorillo, 2013, Davis et al., 2013, Tranos et al., 2015, Danchev and Porter, 2018, Windzio, 2018). In the present chapter, we focus on macroscale migration networks and especially on the global migration network. See Figure 1 for an illustration of a global migration network, with a set of nodes that encode countries and a set of edges that encode migration ties between them.

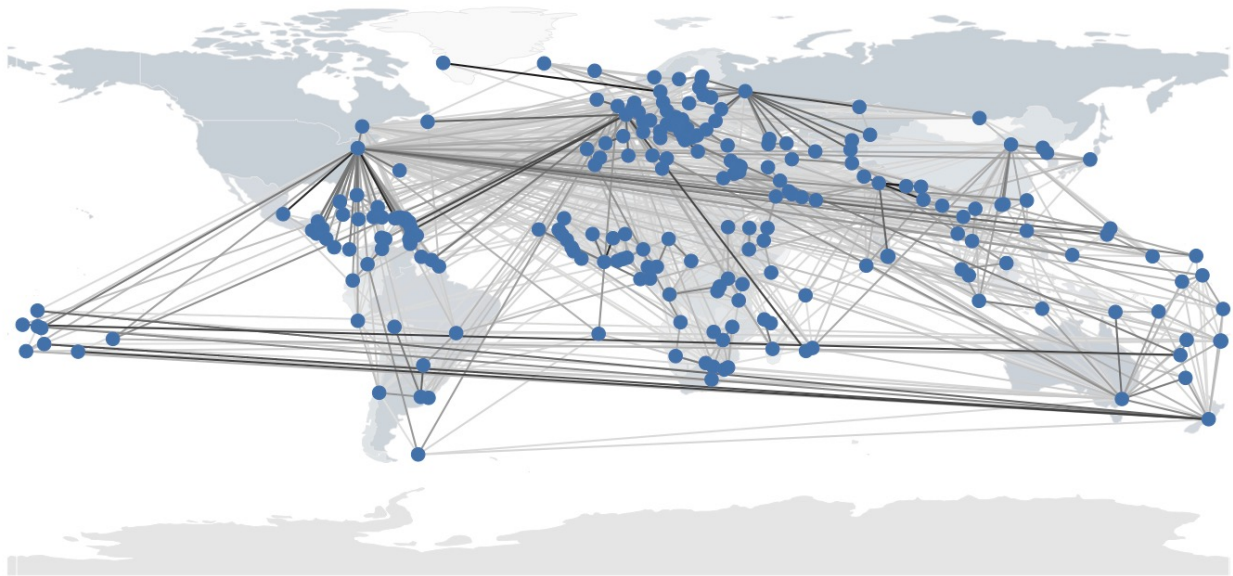

**Figure 1.** An example of a global migration network between countries. The positions of the nodes indicate the geographical locations of the countries. The edges (which are sometimes called "ties" or "links") encode bilateral migrant stock (Özden et al., 2011). We represent population size in the map by varying the gray scale (where darker tones indicate larger populations). [We created the network visualization using code from Traud et al. (2009) and Jeub et al. (2015) in MATLAB and the world map in the background using the package 'rworldmap' in R (South, 2011).] A similar figure appears in Danchev (2015) and in Danchev and Porter (2018).

Until recently, networks were used in migration theory and research (Kritz and Zlotnik, 1992, Salt, 1989) primarily as a metaphor and without the application of explicit network-based methodology. Since the 2010s, however, various studies have employed concepts and methods from network analysis to study migration. A recent special issue in the journal *Social Networks* (Bilecen et al., 2018) included research at the intersection of social networks and migration. Apart from the dyadic assumptions that underlie many migration theories (Massey et al., 1998), a major barrier to network analysis of macroscale international-migration patterns was the lack of compatible migration data between world countries (Fagiolo and Mastrorillo, 2013, Özden et al., 2011). However, over the last decade or so, researchers have had access to public global and regional longitudinal migration data from national statistics and administrative sources, including global bilateral migrant stock (Özden et al., 2011, UN DESA, 2019) and estimates of global bilateral migration flows (Abel and Sander, 2014, Abel, 2018). An important limitation of these migration data sets is the aggregate nature of the information that they provide, although recent data sets stratify migration by different characteristics, including age and sex (UN DESA, 2019, Abel, 2018). Computational and geospatial techniques have been used to infer migration trajectories from individual-level digital geolocated data from online platforms, such as Twitter (Zagheni et al., 2014) and Facebook (Spyratos et al., 2019). As we will discuss in our section on *Future Research Directions: Digital Technologies and Migration Networks*, digital geolocated data can provide an opportunity for increasingly realistic network analyses of migration networks, but they have been used only rarely in this context (Messias et al., 2016). Moreover, to facilitate future research opportunities, it is important to overcome several methodological, privacy, and proprietary limitations of digital geolocated data.

## Migration Networks: Directed, Weighted, Temporal, and Spatial Properties

Danchev and Porter (2018) studied world migration in the form of a directed, weighted, temporal, and spatial network. In this section, we discuss these four features of world migration networks.

As Ravenstein observed long ago (1885), when thinking about migration, one needs to consider the directions of movements, as movement from place A to place B is distinct from movement in the opposite direction. To encode migration direction, one can construct a directed migration network, in which each edge has an associated direction that corresponds to either out-migration or in-migration. Migration movements also vary in terms of volume of migrant stock or flows. One can represent migration volumes by constructing a weighted network (Newman, 2018), in which each edge has an associated weight that encodes the flow or volume of migrant stock from place A to place B for all origin–destination pairs (A,B). Migration networks are also longitudinal, as reflected in global bilateral migration databases, which typically report migrant stock and flows at five-year or ten-year intervals. Rather than considering those time-windows as separate snapshots, recent network methods (Holme and Saramäki, 2012, Holme and Saramäki, 2019) and statistical models (Krivitsky and Handcock, 2014) enable increasingly realistic investigations of network dynamics.

Migration networks are also spatial in nature (Danchev and Porter, 2018, Salt, 2013 [1986]). Spatial networks (Barthelemy, 2018) include both spatially-embedded networks, whose nodes and the edges are embedded in space in a literal sense (e.g., road networks), and networks that are "merely" influenced by space, in the sense that space affects the probability of edge formation and/or the weights of the edges. Migration networks fall into the second category. Because of geographical constraints (Barthelemy, 2018), longer-distance migration edges are associated with costs, so they are less likely than shorter-distance edges to form and develop into strong edges[3] Early migration theories treated the effects of distance on migration as a "law" (Ravenstein, 1885, Zipf, 1946). However, because of significant advances in transportation, information, and communication technologies, geographical and cultural distances have shrunk since the 1970s in a phenomenon that is sometimes called "distance-shrinking" or "time–space compression" (Brunn and Leinbach, 1991, Harvey, 1989). Distance-shrinking technologies have increased the length and the spread of migratory movements (International Organization for Migration, 2003: 16, Vertovec, 2010), although restrictive border control and global inequalities have led to distance-shrinking having smaller effects on migration than it has on other types of cross-border exchanges (such as capital and goods) (Hatton and Williamson, 2005). The effects of distance on migration varies across the world, and it is important to take this spatial heterogeneity into account in network models and computations (Danchev, 2015, Danchev and Porter, 2018).

---

[3] Some networks of human mobility have more stringent spatial constraints than migration networks. For example, daily commuting (Montis et al., 2013) depends on spatially-embedded networks (e.g., transportation systems).

# Methods for Investigating Migration Networks

We now discuss some network diagnostics, techniques, and models that are useful for characterizing the structure of migration networks.

## *Network Diagnostics*

Many diagnostics have been developed to measure the properties of nodes, edges, and other network structures (Newman, 2018; Wasserman and Faust, 1994).

Node degree is perhaps the simple diagnostic for characterizing a country in a migration network. A node's degree is equal to the number of edges that are attached to it. In directed migration networks, one distinguishes a country's out-degree (i.e., the number of outgoing edges that originate at a node) from its in-degree (i.e., the number of incoming edges that terminate at a node) (Wasserman and Faust, 1994: 126). Out-degree and in-degree correspond to out-migration and in-migration, respectively. In weighted networks, it is also useful to consider a node's strength, which one can quantify as the total weight of the edges that are attached to a node (Barrat et al., 2004: 2).

Node degree and node strength are basic measures of "centrality" in networks. Centrality measures are ways to quantify the importances of nodes, edges, or other structures in a network (Newman, 2018). In network analysis, out-degree can be an indication of expansiveness, whereas in-degree is often a notion of popularity (Wasserman and Faust, 1994, Opsahl et al., 2010). From this perspective, one expects that potential migrants from a country with a large out-degree have more opportunities (in the form of potential destinations). Nodes that have disproportionately more connections than the other nodes in a network are sometimes called "hubs" (Newman, 2018). In migration networks, hub countries are involved in the circulation of migrants to and from multiple countries. For discussions of other centrality measures (e.g., betweenness, closeness, and eigenvector centralities), see Borgatti et al. (2018) and (Newman, 2018). We also note that one can evaluate the heterogeneity of centrality scores across nodes in a network by calculating network centralization (Freeman, 1978, Borgatti et al., 2018).

An old idea in migration studies is that "every migratory current has a counter-current" (Grigg, 1977: 112, Ravenstein, 1885: 199). To capture this intuition, consider the network notion of "reciprocity", which measures if an edge from node A to node B is matched by an edge in the opposite direction. One way to define reciprocity is as the number of pairs of mutually connected nodes (i.e., reciprocal relationships) divided by the total number of node pairs with any edge between them (Butts, 2008: 27, Borgatti et al., 2018, Hanneman and Riddle, 2011). One way to generalize reciprocity to weighted networks is to consider two weighted edges as reciprocated if the ratio between them is at least a certain threshold.

A characteristic property of many social and spatial networks is a tendency towards triadic closure (Davis, 1967, Wasserman and Faust, 1994, Barthelemy, 2018). In colloquial sociological terms, triadic closure refers to the tendency of friends of a node to themselves be friends. More formally, triadic closure refers to the probability that two nodes with a connection to a common third node are themselves connected directly to each other (i.e., "adjacent") via an edge (Wasserman and Faust, 1994, Newman, 2018). Both geographical proximity and social proximity (e.g., common language) facilitate triadic closure. Measures of the tendency of triadic closure in unweighted (i.e., binary) networks include clustering coefficients (Newman, 2018) and transitivity (Wasserman and Faust, 1994). There are also generalizations of clustering coefficients to directed (Fagiolo, 2007), weighted (Onnela et al., 2005, Saramäki et al., 2007), and multiplex (Cozzo et al., 2015) networks.

Reciprocity, clustering coefficients, and related measures help characterize node neighborhoods, but they provide limited information about the overall structure of a network. One can use the mean shortest-path length, a measure of network distance between two nodes, as one indicator of global connectivity in a network. A path in a network is a sequence of adjacent nodes, so one can 'travel' from one node to another along these edges. The length of a path in an unweighted network is equal to the number of edges that are traversed. A shortest path is a path that connects two nodes using the fewest possible number of edges (Newman, 2018: 132). In a directed network, one considers paths between origin nodes and destination nodes, and one then calculates a path length by counting the number of edges from an origin to a destination. In weighted networks, one can also transform from edge weights to edge costs to examine other notions of path lengths.

In the study of networks, it is also helpful to compute densities. The most common type of network density is edge density, which is the ratio of the actual edges in a network to the maximum possible number of edges in the network (Wasserman and Faust, 1994: 129, Borgatti et al., 2018). It takes values between 0 (if no edge is present) and 1 (if all edges are present).

## Spatial Network Diagnostics

To examine spatial properties of migration networks, one can calculate diagnostics that combine network and geographical information. For example, a simple way of incorporating spatial information is to compute the probability of a migration edge in a given distance range, taking into account both actual and possible migration edges for that distance range. For a similar approach in the context of the "geography" of online social networks, see Backstrom et al. (2010). See Barthelemy (2018) for a review of spatial networks.

## Community Detection

The systems approach to international migration defines a "migration system" as a set of countries with close historical, cultural, and economic linkages that exchange large numbers of migrants (Kritz et al., 1992, Fawcett, 1989, Salt, 1989). A major methodological difficulty is the demarcation of the boundaries of migration systems (Zlotnik, 1992). Techniques of community detection offer a family of algorithmic methods to delineate migration systems and other "functional regions" (Ratti et al., 2010, Farmer and Fotheringham, 2011) on the basis of empirical connectivity. A "community" is a tightly-knit subnetwork of densely connected nodes that are loosely connected to the rest of a network (Porter et al., 2009, Fortunato and Hric, 2016). In Danchev and Porter (2018), we defined an international-migration community as "a tightly-knit group of countries with dense internal migration connections […] but sparse connections to and from other countries in a network".

There has been an enormous proliferation of methods for algorithmic community detection in networks (Porter et al., 2009, Fortunato and Hric, 2016). One popular method, which has been employed in many network studies of global migration (e.g., Fagiolo and Mastrorillo, 2013, Davis et al., 2013, Tranos et al., 2015), is to maximize an objective function called "modularity" (Newman and Girvan, 2004, Newman, 2018). From a modularity-maximization perspective, an optimal division of a network into communities is one with the largest possible number (or total weights, in a weighted network) of intra-community edges compared to the expected number of such edges in a specified null model (Newman, 2006b, Bassett et al., 2013: 2). The purpose of a null model is to take into account 'statistically surprising' connectivity (Newman, 2006b: 8578).

The standard null model for modularity maximization (Newman, 2006a) works for (either unweighted or weighted) undirected networks. For migration networks, one can consider extensions of modularity that accommodate edge directionality and node attributes. For example, Leicht and Newman (2008) developed a null model for directed networks. Additionally, Expert et al. (2011) and Sarzynska et al. (2016) developed null models for spatial networks (with known node locations), and Mucha et al. (2010) extended modularity maximization to time-dependent and multiplex networks. The choice of community-detection method depends on the properties of available migration data and research questions. When performing community detection, it is important to consider the parameter space, assumptions, and features of a method. For a recent review of community detection (including discussions of increasingly popular methods based on statistical inference), see Fortunato and Hric (2016).

### *Statistical Network Models*

For testing network hypotheses, one can employ statistical models for social networks. The quadratic assignment procedure (QAP) of regression is appropriate for testing dyadic hypotheses (Dekker et al., 2007). Additionally, a popular family of models called exponential random graph models (ERGMs) allows both cross-sectional and longitudinal examination of higher-order network dependencies, and it accounts for covariates that are encoded in node attributes. For a review of ERGMs, see Lusher et al. (2013). For a discussion of various statistical models for network data, see Kolaczyk (2009).

# Prior Research and Key Findings on Migration Networks

Probably one of the earliest works to integrate network analysis and migration studies is a paper by Vincent and Macleod (1974), who made analogies with physical networks (such as stream networks) to advance the argument that networks can influence migration patterns and therefore can inform the forecasting of migration rates. Vincent and Macleod (1974) examined patterns of internal migration by drawing on theories in regional science and on network methods in geography (Haggett and Chorley, 1969). In a pioneering study, Nogle (1994) used the systems approach to international migration, which was proposed by Fawcett and Arnold (1987) and Kritz and Zlotnik (1992), as a framework for studying migration flows within the European Union in the 1980s. By calculating centrality measures and applying techniques for detecting fully connected subgraphs (so-called "cliques"), Nogle (1994) identified a tendency towards a 'Single Europe', in which more countries become interconnected via migration over time. More recently, Maier and Vyborny (2008) applied network analysis in an exploratory study of internal migration between states in the United States. An important contribution of theirs was to define migration as a "mechanism that connects 'places'" and to differentiate the analysis of migration patterns from individual-level analysis of migration decisions and motives of migrants (Maier and Vyborny, 2008). Slater (2008) also examined a network of internal migration in the United States, with a focus on the role of "hubs" and "functional regions". Lemercier and Rosental (2010), inspired by the approach of Hägerstrand (1957) on migration fields, designed an innovative study of migration patterns between rural areas in 19[th] century Northern France using an actor-oriented model for network dynamics (Snijders et al., 2010).

In the last few years, many studies have employed network approaches to study global migration (Tranos et al., 2015, Davis et al., 2013, Fagiolo and Mastrorillo, 2013, Danchev, 2015, Novotný and Hasman, 2016, Abel et al., 2016, Peres et al., 2016, Danchev and Porter, 2018, Windzio, 2018, Cerqueti et al., 2019). By calculating various network diagnostics and employing community detection, several papers (e.g., Tranos et al., 2012, Davis et al., 2013, Fagiolo and Mastrorillo, 2013) highlighted stylized observations about the network of international migration in the latter half of twentieth century. For example, Davis et al. (2013) and Fagiolo and Mastrorillo (2013) concluded that interconnectivity and globalization of migration have increased over time based on increasing "connectivity" (as reflected by the increase in value of various network diagnostics, such the number of migration edges and migration weights between countries, countries' degrees and strengths, and clustering coefficients) and "reachability" (as reflected by decreasing mean shortest-path length). Davis et al. (2013) and Fagiolo and Mastrorillo (2013) also reported that these migration networks have a characteristic right-skewed edge-weight distribution, indicating that many edges have a small to moderate number of migrants and a small number of edges are responsible for many migrants. Additionally, Fagiolo and Mastrorillo (2013: 4) reported that "[t]he number of communities decreases across time" and concluded on this basis that "globalization has made the architecture of the IMN [International Migration Network] less fragmented and modules more strongly interconnected between them". Similarly, Davis et al. (2013: 6) argued their case of "increasing globalization" by noting that "the ratio between the internal and total fluxes slowly decreases in time: 0.8 in 1960; 0.8 in 1970; 0.76 in 1980; 0.75 in 1990 and 2000." A major conclusion of these studies is that world migration has become more interconnected, in line with broader globalization tendencies.

Informed by the international-migration systems approach (Salt, 1989, Kritz et al., 1992), some research (DeWaard et al., 2012, Abel et al., 2016) has employed community detection and other techniques to discover and characterize boundaries of migration systems. Abel et al. (2016) suggested that global migration in the second half of the 20th century is divided into "geographically concentrated" systems that bound neighboring countries in geographical regions. This line of research, which lies at the intersection of migration studies and geography, has emphasized localizing spatial tendencies in international migration.

Many of the above results rely on the structure and boundaries of migration communities that were obtained by maximizing the original modularity function (Newman and Girvan, 2004) or employing some other community-detection algorithm (Fortunato and Hric, 2016) that was designed for non-spatial networks. In a recent paper (Danchev and Porter, 2018), we employed a spatial modularity function (Expert et al., 2011, Sarzynska et al., 2016) that incorporates the probability that a set of countries are assigned to the same community as a function both of migration ties and of the distance between them. In Danchev and Porter (2018), we detected international migration communities using a generalized modularity function for spatial, temporal, directed, and weighted networks. We also leveraged properties of the detected communities to adjudicate among conflicting theoretical accounts. We concluded that, over the second half of the 20th century, "world migration is neither regionally concentrated nor globally interconnected, but instead exhibits a heterogeneous connectivity pattern that channels unequal migration opportunities across the world." Given appropriate data availability and quality, tailored community-detection techniques have the potential to help uncover the impact of recent events— including the enlargement of the European Union, the global financial crisis, and the withdrawal of the United Kingdom from the European Union ("Brexit")—on patterns of connectivity in global migration.

Windzio (2018) employed cross-sectional and longitudinal ERGMs to examine the impact of various factors—including geographical, demographic, economic, and linguistic ones—on migration between 202 countries during the period 1990–2013. They reported results that are consistent with migration theories, with (1) geographical distance tending to reduce the probability of migration edges between countries and (2) economic differences, shared geographical region, and similar religion and language tending to increase the probability of migration edges. Although they fit models using primarily binary data, the ERGM analysis of Windzio (2018) provides a good foundation for sophisticated modeling of network dependencies and countries' attributes in global migration.

## Future Research Directions: Digital Technologies and Migration Networks

Recent advances in digital technologies have created new opportunities to generate, store, and analyze unprecedented amount of online geolocated data from the World Wide Web, social media, and mobile-phone networks (Lazer et al., 2009, Salganik, 2018). When online geolocated data are publicly available and accessible, they provide a valuable opportunity to responsibly study social interactions (Lazer et al., 2009, Salganik, 2018) and human mobility (González et al., 2008). In principle, one can exploit the availability of such digital geolocated data, along with computing infrastructure and scalable algorithms, to increase understanding of migration networks. However, the adoption of those digital innovations in research on migration networks has been relatively slow because of various methodological, ethical, data-availability, and reproducibility challenges that are associated with social data from digital sources (Ruths and Pfeffer, 2014, Olteanu et al., 2018, Salganik, 2018, Jasny et al., 2017). In this section, we discuss both the promise and the challenges of digital technologies for studying migration networks.

### *Digital, Geolocated, and 'Big' Migration Data*

For a long time, research on macroscale patterns of international migration has relied exclusively on administrative records and national statistics (such as population censuses, surveys, and population registers). Although such data are useful in many respects (e.g., because of their wide geographical coverage and public availability), they have important limitations (Rango and Vespe, 2017). First, data from administrative records and national statistics are compiled and available only after a substantial time lag (Zagheni et al., 2017, Rango and Vespe, 2017). Second, country-level aggregation may prove to be too coarse spatially for insightful network analysis. Because country-level aggregate data may obscure migration patterns, it is important to employ location-specific data about movements between actual settlements to conduct increasingly thorough and finely-grained analysis of international migration. Consider, for example, that over 95% of the Bangladeshis (estimated at 200,000 people in the mid-1980s) in Great Britain originated from specific villages in the urban area of Sylhet, which is located in the northeastern part of Bangladesh (Gardner, 1995: 2, Skeldon, 2006: 22, Skeldon, 2018).

Online geolocated data can provide granular and timely information about migration. Recent research has studied international-migration trends using non-representative digital geolocated data of human mobility from the World Wide Web, social media, and mobile phones (Rango and Vespe, 2017, Spyratos et al., 2019, Ahas et al., 2016, State et al., 2014, Zagheni et al., 2014, Böhme et al., 2020). Some studies have inferred geolocated information about migration from geographical (longitude and latitude) coordinates associated with data from Twitter (Zagheni et al., 2014), Google Trends (Böhme et al., 2020), IP (Internet Protocol) addresses (Zagheni and Weber, 2012), and mobile-phone networks (Ahas et al., 2016). Other studies have used anonymized and publicly available self-reported geolocated data, such as country of employment in the professional-networking platform LinkedIn (State et al., 2014), previous locations of residence in Facebook's advertising platform (Spyratos et al., 2019), and geotagged photographs in the photograph-sharing platform Flickr (Barchiesi et al., 2015).

Although not without limitations, which we detail later, digital geolocated data from social media is a promising resource that one can extend to the study of migration networks. For example, Spyratos et al. (2019, see also Zagheni et al., 2017) used publicly available and aggregate information from Facebook's advertising platform to estimate stocks of migrants—specifically, using the number of Facebook users in each country who previously resided in a different country—for 119 destination countries. In principle, one can apply network analysis to such data (and similar data of migrant stocks from other social-media platforms), but one also has to somehow address the problem of missing nodes and edges (e.g., from heterogeneous and potentially limited coverage of countries). To take such limitations into account, it will be helpful to use methods for inferring missing nodes and edges in networks (Kossinets, 2006, Guimerà and Sales-Pardo, 2009, Kim and Leskovec, 2011). One also needs to consider other likely limitations— including selection biases and undocumented criteria for establishing previous country of residence (Spyratos et al., 2019)—of data from Facebook and other social-media platforms.

Another promising research avenue is to leverage data from online activities (e.g., Web searches) to anticipate future movements in the context of broader patterns in migration networks. For example, Böhme et al. (2020) combined geolocated online search data from Google Trends with bilateral migration flows from the OECD International Migration Database (IMD) and survey data on migration intentions from the Gallup World Poll (Tortora et al., 2010) to estimate international-migration flows. Along those lines, albeit at a smaller scale, a Pew Research Center report (Connor, 2017) combined traditional migration data with geolocated data of Web searches to track migration patterns of refugees from Middle East to Europe. These studies did not use network analysis, but (to the extent that the selection of any destination depends in part on other destinations) one can construct migration networks and track their dynamics to anticipate changes in some of the countries of origin and destination, migration ties, and other structures (such as migration communities).

Data sources from mobile-phone call-detail records (CDRs) provide another opportunity to harness new technologies to track patterns of migration (Rango and Vespe, 2017) and thereby construct migration networks. One example is the Data for Development (D4D) challenge that released anonymized CDRs (de Montjoye et al., 2014), which were then used to study networks of internal-migration patterns (Martin-Gutierrez et al., 2016). One can also combine CDR data with census data. For example, Eagle et al. (2010) linked communication networks with national census data to study associations between network structure and socio-economic opportunities. Anonymized geolocated data from CDR have typically been useful for tracking internal-migration patterns in a single country (Rango and Vespe, 2017) although studies have used both domestic and roaming CDRs to construct star-like networks and study "transnational" mobility between a seed country and other countries (Ahas et al., 2016).

Despite their immense potential promise, it is important to acknowledge the limitations of data from social-media platforms, CDRs, and other digital sources for the study of migration networks (International Organization for Migration's Global Migration Data Analysis Centre (IOM's GMDAC), 2017). In addition to the fact that geolocated data from online sources are typically non-representative, which is widely acknowledged, such data are often more restricted in their geographical scope than is the case in global migration databases. Importantly, because digital geolocated data typically include traces of individual human behavior, protection of privacy and confidentiality is a major concern (De Montjoye et al., 2013, Rocher et al., 2019). The problem is particularly pertinent to network data because  dependencies between observations (a hallmark of network data) complicate efforts at anonymization. Furthermore, digital geolocated data are often proprietary and rarely available to others to openly reuse and/or reanalyze, making it impossible to evaluate both the computational workflow and the reproducibility of results (Stodden et al., 2016, Jasny et al., 2017). Recent initiatives, such as D4D-Senegal (de Montjoye et al., 2014) and the Social Science One program (King and Persily, 2019), have offered secure solutions for leveraging proprietary and individual-level data for open research while simultaneously preserving privacy and data integrity. The success of such socio-technological innovations of data access and reuse can transform computational research of both migration networks and other areas of human science.

### *Advances in Network Analysis and Related Methodologies*

Prior research on migration networks has largely employed standard techniques and diagnostics from network analysis. However, the analysis of complex and "big" geolocated data on migration networks requires the application and development of new computational methods and models. Examples of recent advances in network methodologies, which we expect to be valuable for research on migration networks, include work on machine learning and networks (Grover and Leskovec, 2016), new spatial null models for community detection (Sarzynska et al., 2016), Bayesian network models (Peixoto, 2019), multilayer network analysis (Kivelä et al., 2014), and spatial applications of topological data analysis (Feng and Porter, in press (2020)).

### *Multilayer Networks of Migration and Other Exchanges Between Places*

In this chapter, we have restricted our discussion thus far to ordinary ("monolayer") networks, which include a single type of edge (e.g., with one value to encode bilateral migration stocks or flows). However, since the late 1980s, international migration has been viewed in the context of other spatial interactions, including economic ones (e.g., international trade), historical ones (e.g., previous colonial ties), cultural ones (e.g., language), and political ones (e.g., bilateral agreements) (Malmberg, 1997: 40, Kritz and Zlotnik, 1992, Portes and Böröcz, 1989, Fawcett, 1989). Additionally, research using data from social-media platforms indicates that there is a strong association between international social relationships and international migration (Takhteyev et al., 2012, Hale, 2014, Chi et al., 2020). To encode those various relationships between places, one can construct a multiplex network (Wasserman and Faust, 1994, Kivelä et al., 2014, Aleta and Moreno, 2019), in which nodes can be linked to each other via more than one type of relationship. A multiplex network has multiple "layers", which in the above example encode economic, historical, cultural, political, or communication relationships between countries. There has been little empirical examination of the relationships between multiple layers of international relationships, apart from Belyi et al. (2017) and network research on migration and trade (Sgrignoli et al., 2015, Fagiolo and Mastrorillo, 2014).

Multiplex networks can also facilitate examination of multiple types of international migration. For example, different layers in a multiplex network can encode different policy categories by entry (e.g., work, free movements, family, and humanitarian (OECD, 2016)) or types of migrants (including asylum seekers, refuges, family members, low-skilled labor migrants, students, and high-skilled migrants). Unfortunately, current estimates of migration stocks and flows are either aggregated or stratified by general migrant attributes, such as age (UN DESA, 2019, Brücker et al., 2013), gender (Özden et al., 2011, Abel, 2018), and educational attainment (Brücker et al., 2013, Docquier et al., 2009). Data on types of migrants are scattered and rarely available, except for OECD (Organisation for Economic Co-operation and Development) countries. For some types of migrants, it is possible to use estimates from digital geolocated data. For example, State et al. (2014) examined data from LinkedIn to track international-migration flows of highly-skilled migrants. The construction of a multiplex migration network, however, requires comparable data from additional sources, such as of multiple types of migrants or from multiple social-media platforms. Therefore, it is necessary to integrate multiple data sources, such as ones that are indicative of different types of migration, and use them to construct and analyze multiplex migration networks.

With the availability of harmonized migration data of the types of migrants, multilayer network analysis may provide both a theoretical framework and methodology for systematic investigations of multiple socio-economic relationships and/or multiple types of migration between world countries.

### *Relevance to Theory, Social Issues, and Policy*

To establish its relevance and impact, network-based research on macroscale migration patterns should target research questions of theoretical, societal, and policy importance. Research on macroscale migration networks thus far has primarily tested propositions from prior migration theories (e.g., Windzio, 2018, Danchev and Porter, 2018). Future research efforts should explore opportunities to develop new theoretical propositions. One important theoretical development is to conceptualize the roles of migration networks in perpetuating, generating, or alleviating global inequalities in migration. For example, research on interpersonal networks has explored the conditions under which ethnic migrant networks have beneficial or adverse effects for economic integration of migrants (Portes, 1998, Martén et al., 2019).

Furthermore, to produce policy-relevant findings, research on migration networks should engage with concepts, categories, and legal-policy frameworks at subnational, national, and international levels (Betts, 2011, International Organization for Migration, 2019). One example of policy-relevant work is a recent paper by Bansak et al. (2018), who employed data-driven matching algorithms to assign refugees to jobs to improve refugee integration.

# Conclusions

Advances in transportation, information, and communication technologies have combined with policy, economic, and social forces to reshape migration since the 1970s. The understanding of complex migration patterns cannot rely solely on bilateral-level investigations, and systematic network analysis can help shed light on unknown and unappreciated migration dynamics. For many decades, migration theories (Hägerstrand, 1957, Kritz and Zlotnik, 1992, Zlotnik, 1992, Mabogunje, 1970, Salt, 1989, Fawcett, 1989) developed the intuition of global migration as a network of places that are linked by movements, but these investigations lacked the formal language and techniques of network analysis to systematically encode migration connections between places in a network and study them explicitly from a network-based perspective. In this chapter, we reviewed formalism and techniques from network analysis that have been applied to international migration and have yielded fascinating insights into the structure and dynamics of world migration.

We then discussed advances of digital technologies that have made it possible to generate, store, and analyze large quantities of geolocated data from online sources, mobile phones, and other resources. We outlined ways in which one can combine data from these new sources with more traditional data from national statistics, administrative records, and surveys to study the structure and dynamics of macroscale migration networks. There are many important challenges—including ones that are related to methodology, data privacy, geographical coverage, data availability, and research transparency—that have impacted the slow uptake of new data sources and data integration into the study of global migration networks.

For data sources to be available for reuse (Wilkinson et al., 2016), it is crucial to implement socio-technological innovations to simultaneously improve privacy protection, secure use of proprietary data for research purposes, and promote research transparency and reproducibility. Making advances on issues such as data ethics and data availability will help investigations of migration networks to make better use of new developments in network analysis (e.g., new methods for community detection, multilayer network analysis, and statistical data analysis) and other promising areas of research, including machine learning, Bayesian statistics, and topological data analysis. Finally, it is important for theory-driven and problem-driven investigations that foster public debate and inform evidence-based policy to help guide future research efforts. Leveraging technological advances to make progress towards this goal requires interdisciplinary collaborations across many areas of research, including migration studies, geography, network science, computational social science, digital technologies, machine learning, and mathematics.

## Acknowledgements

We thank Marie McAuliffe for the invitation to write this chapter and for helpful comments. The chapter draws on our prior work on these topics. We thank Marya Bazzi, Adam Dennett, Bernie Hogan, Lucas Jeub, Michael Keith, Mikko Kivelä, Renaud Lambiotte, Martin Ruhs, and Marta Sarzynska for helpful discussions.